\documentstyle[preprint,cite,aps]{revtex}
\newcommand{\be}{\begin{equation}}
\newcommand{\ee}{\end{equation}}
\newcommand{\bea}{\begin{eqnarray}}
\newcommand{\eea}{\end{eqnarray}}
\newcommand{\ba}{\begin{array}}
\newcommand{\ea}{\end{array}}

\newcommand{\bb}{\bibitem}

\begin{document}
\draft
\tightenlines

\title{\bf Reply to ``Comment on Renormalization group picture of the
Lifshitz critical behavior''}
\author{Marcelo M. Leite\footnote{e-mail:leite@fis.ita.br}}
\address{{\it Departamento de F\'\i sica, Instituto
Tecnol\'ogico de Aeron\'autica, Centro T\'ecnico Aeroespacial,
12228-900, S\~ao Jos\'e dos Campos, SP, Brazil}}
\maketitle

\vspace{0.2cm}
\begin{abstract}
{\it We reply to a recent comment by Diehl and Shpot (cond-mat/0305131)
criticizing a new approach to the Lifshitz critical behavior just presented
(M. M. Leite, Phys. Rev. B {\bf 67}, 104415(2003)). We show that this approach
is free of inconsistencies in the ultraviolet regime. We recall
that the orthogonal approximation employed to solve arbitrary loop diagrams
worked out at the criticized paper even at three-loop level are consistent
with homogeneity for arbitrary loop momenta. We show that the criticism
is incorrect.}
\end{abstract}

\vspace{1cm}

\newpage

Diehl and Shpot (DS) \cite{DS1} recently formulated a criticism of a new
renormalization-group(RG) picture of the Lifshitz critical behavior
\cite{L1}. It is based on a scaling hypothesis with two independent
relevant length(momentum) scales, which characterize the spatial axes
without competition as well as those competing space directions
\cite{L2}. In momentum space, the Feynman diagrams are calculated up
to two-loop order using two different approximations. The theory is
renormalized using dimensional regularization in two different
renormalization schemes. We first use normalization conditions with
two distinct symmetry points characterizing different momenta
directions. Then, we check our results using a minimal subtraction
scheme.

The dissipative approximation was used to calculate some critical
exponents along directions perpendicular to the (quartic) competing
directions \cite{AL1}. The main point of the criticism by DS in
\cite{DS2} to this approximation was the impossibility to treat the
isotropic case. However, it was pointed out in \cite{AL2} that it is
a good approximation for the anisotropic behaviors, since it preserves
the homogeneity of the Feynman integrals in the external quadratic
momenta components perpendicular to the competition axes.

The orthogonal approximation to perform loop integrals introduced
in \cite{L1} is the most general one consistent with the physical
principle of homogeneity. It can address both isotropic and
anisotropic cases since the loop integrals are homogeneous functions
of arbitrary external momenta scales perpendicular to or along the
competing axes. Therefore, the main point of DS in \cite{DS2} no
longer apply for the orthogonal approximation presented in \cite{L1}.

The criticism in \cite{DS1} has a different nature: the authors claim that
``(i)Leite's renormalization scheme does not yield an ultraviolet finite
renormalized theory, and the structure of the RG he formulates is incorrect.''
Let us show now why this statement is wrong. Recall that each
vertex part in Ref.\cite{L1} has a subscript $\tau =1,2$
(see for example Eq.(6)). If $\bf{k}$ is a vector along the $m$ competing
directions and $\bf{p}$ is a vector along the noncompeting $(d-m)$
directions, the vector $\bf{q} = (\bf{k}, \bf{p})$ is the most general
$d$-dimensional momentum. When $\tau=1$, the vertex part
$\Gamma_{R (1)}^{(2,0)}(\bf{q})$ has nonvanishing external momenta
components only along directions perpendicular to the competing axes, i.e.,
$\bf{q} = (0, \bf{p})$ . The associated renormalization factors
$Z_{\phi (1)}, Z_{\phi^{2} (1)}$ and renormalized coupling constant
$u_{1}$ are defined through $(2a)-(2e)$ in such a way that the renormalized
vertex part with $\Gamma_{R (1)}^{(2,0)}(\bf{p})$ (given by Eq. (193a) for
$\tau=1$) in \cite{L1} is ultraviolet finite. When $\tau=2$, the vertex part
$\Gamma_{R (2)}^{(2,0)}(\bf{q})$ has nonvanishing external momenta only
along directions parallel to the competing axes, i.e.,
$\bf{q} = (\bf{k}, 0)$. {\it The authors miss that
in addition} one has the renormalization factors
 $Z_{\phi (2)}, Z_{\phi^{2} (2)}$ and renormalized coupling constant
$u_{2}$ which are defined through $(3a)-(3e)$ with renormalized vertex
$\Gamma_{R (2)}^{(2,0)}(\bf{k})$ (given by Eq. (193a) for
$\tau=2$) which is ultraviolet finite. In minimal subtraction,
for $\tau=1$, the functions $Z_{\phi (1)}$, $Z_{\phi^{2} (1)}$ and
$u_{1}$ are defined in Eqs. (192a)-(192c). Eq.(193a) defining the
renormalized vertex $\Gamma_{R (1)}^{(2,0)}(\bf{p})$, and Eqs.(194a),
(195d) expressing the bare vertex $\Gamma_{(1)}^{(2,0)}(\bf{p})$ eliminates
the ultraviolet pole proportional to $\frac{p^{2}}{\epsilon_{L}}$,
making the renormalized vertex $\Gamma_{R (1)}^{(2,0)}(\bf{p})$ ultraviolet
finite as shown explicitly there. The pole proportional to
$\frac{k^{4}}{\epsilon_{L}}$ of the bare vertex
$\Gamma_{(2)}^{(2,0)}(\bf{k})$ is explicitly eliminated in VIB2 using a
similar reasoning with $\tau=2$. These explicit cancellations
in this minimal subtraction scheme first appeared in Ref.\cite{L1}. Indeed,
the minimal subtraction was carried out up to three-loop level for
$\Gamma_{R (\tau)}^{(2,0)}(\bf{q})$.

Notice that if one tries to renormalize the theory using minimal
subtraction using the vertex $\Gamma_{R}^{(N,L)}( \bf{k}, \bf{p})$
with arbitrary momenta, without separating each subspace into independent
RG transformations, the renormalized vertices are not finite in the
ultraviolet regime. Had we not separated the renormalized vertices in
that way we would have obtained a renormalized theory with bad ultraviolet
behavior. This separation is possible, for the two coupling constants
flow consistently to the same fixed point. Given these facts,
the claim (i) is incorrect.

Next, the authors make the claim ``(ii) Leite's insufficient choice of
counterterms is biased towards giving the incorrect value
$\theta=\frac{1}{2}$ for the anisotropy exponent
$\theta=\frac{\nu_{L4}}{\nu_{L2}}$''. As shown in the above paragraph , the
choice of counterterms is not insufficient. Moreover, the critical exponents
$\nu_{L4}$ and $\nu_{L2}$ are determined independently in the perturbative
framework up to two-loop level. The value $\theta=\frac{1}{2}$ is just a
simple consequence of this analysis. In the following discussion of this
claim, the authors insist that the choice of counterterms is insuficcient.
Hence, the claim (ii) is unwarranted.

The following claim is ``(iii) Leite obtained incorrect hyperscaling
relations because he missed the fact that $\theta$ is an independent
exponent, not identical to $\frac{1}{2}$ for all $\epsilon_{L}>0$''.
The hyperscaling relation is derived from the specific heat vertex
part above the critical Lifshitz temperature, and relates the specific heat
exponent with the space dimension and the correlation length exponents, as
stated in Eqs.(45) for the anisotropic cases \cite{L1,Am}. We recall that the
exponent $\beta_{L}$ is obtained when performing the RG analysis {\it below}
the critical Lifshitz temperature, as shown in Eqs.(54b) and (54d). Of course,
since the specific heat critical exponent is the same above and below
$T_{L}$, the magnetization exponent $\beta_{L}$ can be related to
$\alpha_{L}$ and $\gamma_{L}$ through the Rushbrook law.  In
\cite{L1} it was explicitly demonstrated that the anisotropic scaling
relations are identical to that in the seminal paper \cite{HLS}. The
Eqs. (54b) and (54d) do satisfy Eq. (1) in Ref.\cite{DS1} for
$\theta = \frac{1}{2}$, which is the correct value of $\theta$, at least
at two-loop level. From our scaling analysis $\theta$ is not an independent
exponent. Therefore, claim (iii) is out of order.

The claim (iv) is about the role of $\sigma$. As shown in the text \cite{L1},
$\sigma$ is not required, since we develop two independent sets of
normalization conditions in each subspace. Furthermore, if $\sigma$ is set
to unit and the external momenta along the competing axes have the same
canonical dimension as the components perpendicular to the competing axes,
the quartic kinetic term in the Lagrangian accounting for the effect of the
competition is inconsistent for it has the wrong canonical dimension (in
mass units). This invalidates (iv).

The claim (v) says that the results obtained in Ref. \cite{L1} for the
isotropic case are false. Let us first analyse the scaling
laws. The scaling laws in \cite{L1} are identical to those obtained in the
earlier work \cite{Ni} for isotropic cases with arbitrary even
momentum powers $p^{2L}$ in the propagators when $L=2$. It is
important to mention that DS treatment was unable to derive these
scaling laws.

Consider the one-loop Feynman integral $I_{2}(K')$ Eq.(150) from
\cite{L1}. It can be calculated to order $\epsilon_{L}^{0}$ without
any approximation as follows. Using Feynman parameters, Eq.(150)
reads:
\begin{equation}
I_{2}(K') = \Gamma(4) \int_{0}^{1} dx x(1-x) \int \frac{d^{m}k}
{[x k^{2} + (1-x)(k+K')^{2}]^{4}}.
\end{equation}

Using the formula

\begin{equation}
\int \frac{d^{m}k}{(k^{2} + 2kk'+ m^{2})^{\alpha}} =
\frac{1}{2} \frac{S_{m} \Gamma(\frac{m}{2}) \Gamma(\alpha -
\frac{m}{2})}{\Gamma(\alpha)} (m^{2} - k'^{2})^{\frac{m}{2} - \alpha},
\end{equation}
we obtain
\begin{equation}
I_{2}(K') = \frac{1}{2} \Gamma(\frac{m}{2}) \Gamma(4 - \frac{m}{2})
S_{m} \int_{0}^{1} dx x(1-x) [x(1-x)K'^{2}]^{\frac{m}{2} -4}.
\end{equation}

The integral above is different from its analogue in the standard
$\phi^{4}$ theory for the appearance of the extra factor $x(1-x)$. By
taking $m=8 - \epsilon_{L}$ and expanding the Gamma functions, we find
up to $\epsilon_{L}^{0}$ with no approximation the result
\begin{equation}
I_{2}(K') = S_{m} 4 \times 3 \times 2 \frac{(1 -
\frac{13 \epsilon_{L}}{24})}{\epsilon_{L}}[ \frac{1}{6} -
\frac{\epsilon_{L}}{2} \int_{0}^{1} dx x(1-x)ln[x(1-x)K'^{2}] +
O(\epsilon_{L}^{2})].
\end{equation}
Notice that the remaining logarithmic integral in the last equation is
momentum dependent. In minimal subtraction this integral does not
need to be calculated. Nevertheless, it has to be considered in order
to show that the renormalization factors are momentum
independent. This condition is achieved provided the cancellations of
all the logarithmic integrals take place for arbitrary vertex parts.
Thus, any attempt to solve the integrals without doing approximations
has to take into account these basic facts.

It is clear from last equation that the remaining logarithmic
integrals in the isotropic case are not the same from those in
the standard $\phi^{4}$ theory. In Ref.\cite{DS4} the validity
of Eq.(A1) for arbitrary external momenta imply that the logarithmic
integrals do not cancell out in the calculation of the renormalization
factors, making them momentum-dependent in contradiction to
Eqs.(12)-(14) in \cite{DS4}. This shows that the results in \cite{DS4}
are inconsistent. On the other hand, the use of the orthogonal
approximation in \cite{L1} provides all the cancellations of
logarithmic integrals for arbitrary vertex parts making the
renormalization factors momentum-independent as explicitly
shown there.

Let us compare our findings for the isotropic case using normalization
conditions. Taking $K'^{2}=1$, the integral above can be easily
calculated giving the result
\begin{equation}
I_{2}(K'^{2}=1) = 4S_{m} \frac{(1 -
\frac{\epsilon_{L}}{24})}{\epsilon_{L}}.
\end{equation}

In \cite{L1} it was incorrectly asserted that a choice of a
convenient factor to be absorbed in the coupling constant would
affect universal quantities. Then, if we choose the factor
$F_{m,\epsilon_{L}}= 4 S_{m}(1 - \frac{7 \epsilon_{L}}{24})$, the
exact result above and approximate form Eq.(157) from \cite{L1} of
$I_{2}(K'^{2}=1)$ are the same and only differ by an ultraviolet finite
reparametrization of the theory. Thus, the orthogonal approximation
for $I_{2}$ is the same as the exact solution up to a finite
ultraviolet reparametrization which does not change universal
ammounts. This invalidates the sentences ``... due to his incorrect
calculation... he gets even the simple one-loop integral $I_{2}(K)$
defined in (150) wrong''. The discussion above implies that (v)
is false. The advantage of the orthogonal approximation is that it
permits to treat the anisotropic and isotropic loop integrals
within the same mathematical footing.

In (vi), the authors actually ``{\it ... fail to see...} $\epsilon$-expansion
results qualify as acceptable approximations.'' In fact Ref.\cite{L1}
achieved the two goals: (a) homogeneity is the physical principle which
justifies the orthogonal approximation; (b) the independent flow of the two
coupling constants along different momenta subspaces to the same fixed point
in the anisotropic cases is consistent and yields a well-defined
theory. The critical exponents and other universal ammounts for the
anisotropic cases \cite{L3,L4} reduce correctly to the cases $m=0$
using this approximation.

DS see no need to make approximations, but there is a point in their
formulation that deserves at least un update to correct a wrong result.
Consider the two-loop integral $I_{4}(\bf{P}, \bf{K'})$ contributing to the
coupling constant at two-loops. Since $\bf{P}$ is a $(d-m)$-dimensional
momentum vector perpendicular to the competing axes and  $\bf{K'}$ is a
momentum vector parallel to the $m$-dimensional competing axes the equations
(148) and (137) for the solution of this integral using the orthogonal
approximation in \cite{L1} depends on {\it both} external momenta.
In Ref.\cite{DS3} Eq. (B.14), the integral $I_{4}(\bf{P}, \bf{K'})$ only
depends on $P$ when ``performing'' the calculations either in momentum space
(as they ``did'' in appendix B) or in coordinate space (see appendix C). This
is obviously incomplete and wrong, since the most general situation should
include both momentum scales. In Ref.\cite{DS4} they tried to defend their
result with a falacious argument in appendix B. The simple orthogonal
approximation presented in \cite{L1} for this integral Eq.(148) simply rules
out Eq. (B.14) in Ref.\cite{DS3} as a valid equation, making it unacceptable.
As was already pointed out in \cite{AL2} the incorrect behavior of this
integral, for instance, prevents the transition from the anisotropic to
the isotropic case.

To summarize, the renormalized field theory in \cite{L1} is free of
ultraviolet pathologies for both the isotropic and anisotropic cases.
DS's misconceptions of the method proposed in that reference lead them
to make a incorrect criticism.

I acknowledge support from FAPESP, grant number 00/06572-6.

\end{document}